М.Л. Калужский

# Трансформация маркетинга в электронной коммерции

*Аннотация*: Статья о трансформации теории и практики маркетинга в условиях электронной коммерции и сетевой экономики. Автор рассматривает Интернет-маркетинг как самостоятельный вид маркетинга в виртуальной коммуникативной среде. Основной тезис статьи: виртуальная среда определяет трансформацию маркетинга, изменяя методы, приоритеты и структуру сначала практики, а затем теории маркетинга.

*Ключевые слова*: электронная коммерция, маркетинг, интернет-маркетинг, комплекс маркетинга.

M.L. Kaluzhsky

# Transformation of marketing in the e-commerce

*Annotation*: The article is about transformation of the theory and practice of marketing in the conditions of e-commerce and network economy. The author considers Internet-marketing as an independent kind of marketing in the virtual communicative environment. The basic thesis of the article: the virtual environment defines marketing transformation, changing methods, priorities and structure at practice and then theories of marketing.

*Keywords*: e-commerce, marketing, Internet-marketing, marketing-mix.

Калужский Михаил Леонидович (Омск, ОмГТУ, frsr@inbox.ru)

Бурное развитие электронной коммерции в последние годы не могло не отразиться на теории и практике продвижения товаров во Всемирной сети. Маркетинг не просто выработал новые приёмы Интернет-торговли. На основе традиционного маркетинга, многократно описанного в учебниках, сформировался т.н. «Интернет-маркетинг», отличительная черта которого заключается в том, что все участники сети находятся в сопоставимо равных стартовых условиях. Интернет-маркетинг имеет ту же структуру, что и традиционный маркетинг, но действует на качественно ином уровне экономических отношений.

Основной сферой приложения усилий Интернет-маркетинга являются трансакционные издержки и новые возможности, связанные с их сокращением. Поэтому первостепенную роль здесь играет не товарная политика (как в традиционном маркетинге), не коммуникативная (как в индустриальной экономике) и даже не маркетинговые исследования (см. табл. 1). Первостепенную роль в Интернет-маркетинге играет сбытовая политика, позволяющая сделать товар доступным для максимального количества потенциальных покупателей.

**Табл. 1. Трансформация комплекса маркетинга в сети Интернет**

| Традиционный маркетинг-микс | Комплекс Интернет-маркетинга |
|---|---|
| 1. Товарная политика | 1. Сбытовая политика |
| 2. Ценовая политика | 2. Ценовая политика |
| 3. Сбытовая политика | 3. Товарная политика |
| 4. Коммуникативная политика | 4. Коммуникативная политика |

Это очень важный тезис, согласно которому *Интернет в сетевой экономике выполняет в первую очередь сбытовые функции*. Субъекты сетевой экономики приходят в Интернет не для коммуникаций или маркетинговых исследований. Они рассматривают Интернет как отдельный большой рынок, на котором существует низкий входной барьер и равные конкурентные возможности. Всё остальное – вторично. Не случайно экономическая деятельность в Интернете получила название «сетевая коммерция», т.е. «*процесс покупки и продажи*» через сеть Интернет.[1]

---

[1] Котлер Ф. Маркетинг менеджмент. – СПб.: Питер, 2009. – С. 781.

Второе место по значимости в Интернет-маркетинге занимает ценовая политика. С одной стороны, это связано с объективным сокращением трансакционных издержек для продавцов, за счёт которого малые виртуальные субъекты могут конкурировать с крупными традиционными участниками рынка. С другой стороны, главным мотивом для совершения покупки в Интернете при условии доступности товара является его сравнительно более низкая цена. Трансакционные издержки в электронной торговле намного ниже, чем в обычной торговле. Поэтому ценовые возможности привлечения покупателей у виртуальных компаний намного больше, чем в традиционном бизнесе.

Только третье место по значимости в Интернет-маркетинге занимает товарная политика. Это обстоятельство обусловлено тем, что виртуальные компании не занимаются производством и продвигают не товары, а информацию о товарах. Электронная коммерция даёт значительно большую мобильность в выборе товаров и поставщиков, чем обычная торговля. В сетевой экономике конкурентное преимущество на рынке создают не товары, а методы их продвижения. Стратегически выигрывает не тот, кто располагает товарами, а тот, кто располагает возможностями их продвижения. Лучший пример – интернет-аукцион «eBay» с выручкой в 2011 году в 11,65 млрд. долларов (рост на 27%) и чистой прибылью в 3,23 млрд. долларов (рост на 79%).[2]

На последнем месте по значимости в Интернет-маркетинге располагается коммуникативная политика так же, как и в традиционном маркетинге-микс.[3] Доказать это тезис очень просто, так как коммуникации не способны продвинуть товар, если товар: а) недоступен, б) дороже аналогов, г) не соответствует ожиданиям потребителей. Этот инструмент работает только тогда, когда отсутствуют проблемы с тремя предыдущими инструментами Интернет-маркетинга.

Инструментарий Интернет-маркетинга, как и инструментарий традиционного маркетинга-микс, укладывается в рамки общеизвестной концепции «4P».[4] Четырёх элементов вполне достаточно, чтобы раскрыть основные направления, инструменты и методы Интернет-маркетинга в условиях сетевой экономики. Отличие заключается лишь в изменении порядка расположения элементов комплекса маркетинга.

**Элемент I. Сбытовая политика в Интернет-маркетинге**. Сбытовая политика в Интернет-маркетинге включает в себя три основных составляющих традиционного маркетинга: обмен и трансакции, отношения между партнёрами и взаимодействие с покупателями.[5] Однако специфика виртуального пространства наполняет их новым, отличным от прежнего, содержанием:

1. *Обмен и трансакции*. Согласно классической теории маркетинга обмен лежит в основе любой коммерческой деятельности. «*Маркетинг появляется в тот момент*, – пишет Ф.Котлер, – *когда люди решают удовлетворять нужды и потребности посредством обмена*».[6] Тогда как под трансакциями в теории маркетинга понимается «*обмен ценностями между двумя и более сторонами*».[7]

Трансакция становится возможной, когда ценности, потребности и интересы участвующих в сделке сторон совпадают. В Интернет-маркетинге базовой ценностью является не товар, а электронные каналы сбыта. Они обеспечивают получение прибыли и являются главным фактором конкурентоспособности за счёт снижения трансакционных издержек.

Именно низкие трансакционные издержки в Интернете позволяют создавать и использовать неограниченное число дёшевых в эксплуатации, круглосуточно работающих в авто-

---

матическом режиме (24-7-365), специализированных каналов сбыта.[8] И пока существует дисбаланс в трансакционных издержках между традиционной торговлей и электронной коммерцией, гарантированная прибыль от использования электронных каналов сбыта будет важнее непостоянной прибыли от традиционных форм маркетинга.

Следствием развития электронной коммерции стало снижение роли традиционной торговой инфраструктуры при осуществлении трансакций. Если значение складской и транспортной инфраструктуры практически не меняется, то торговую инфраструктуру (прилавки, выставочные залы, продавцов и т.д.) с успехом заменяют электронные каталоги и прайс-листы. Поэтому *основная функция сбытовой политики в Интернет-маркетинге подразумевает не создание каналов сбыта, а использование имеющихся в сети возможностей с целью обеспечения присутствия товаров в различных сегментах виртуального рынка*.

Торговая инфраструктура в традиционном маркетинге представляла собой большое количество посредников с задолженностями по поставкам, товарными запасами, спецификой рынков сбыта и связанными с этим проблемами поставщика. Если товар поставлялся по предоплате, то посредники заказывали ограниченное количество товара, требовали увеличения скидок и легко шли на контакт с конкурентами. Если товар отпускался на консигнацию, то происходило затоваривание посредников, замедлялся оборот и начинались проблемы с оплатой поставок. При этом каждый уровень канала сбыта обеспечивал торговую надбавку в размере от минимума рентабельности до бесконечности, в зависимости от удалённости поставщика и степени монополизации географического рынка.

В электронной коммерции расстояния утратили значение, а торговая инфраструктура (оптовое и розничное звенья) выпала из торговой сети.[9] Произошло то, что П.Дойль назвал *«отделением информации от продукта»*, когда посредник имеет дело не с товарами, а с информацией об этих товарах.[10] В результате товарные потоки в каналах сбыта уступили место информационным потокам, а торговля товарами превратилась в информационную поддержку прямых поставок товаров.[11]

Итогом стало привлечение большего количества участников в процесс сбыта товара с одновременным стиранием границ между рекламной деятельностью, потребительским поведением и розничной торговлей. Это позволило Интернет-посредникам обеспечить предложение с максимальным ассортиментом и минимальными издержками.

При условии достаточной логистической поддержки участникам каналов сбыта в Интернет-маркетинге не требуются ни собственные склады, ни торговые площади, ни торговый персонал. Производитель сам или через логистических посредников обеспечивает возможность предложения товара (наличие товара, условия поставок и приём оплаты).

Организацию сбыта берёт на себя сетевое сообщество, которое может состоять как из специализированных торговцев, так и из объединившихся покупателей. При этом формальные институциональные рамки отсутствуют, и переход от роли покупателя к роли продавца может происходить практически мгновенно.

2. *Отношения между партнёрами*. Развитие электронной коммерции привело не только к изменению сбытовой политики в Интернет-маркетинге, но и к изменению характера взаимоотношений участников системы сбыта. Это связано с появлением такого понятия как «е-сорсинг», под которым понимаются *«инструменты, позволяющие выявлять потенциальных поставщиков и в ходе переговоров обговаривать с ними условия, ведущие к самым низким затратам»*.[12] Благодаря е-сорсингу реализация сбытовых функций и распределение заказов окончательно перешли из сферы менеджмента в сферу маркетинга.

---

Трансформация стала настолько глубокой, что потребовала коренного пересмотра содержания и функций некоторых основополагающих инструментов маркетинга. В первую очередь это коснулось рекламы. Электронная торговля вывернула наизнанку саму сущность рекламной деятельности в сети, превратив её из коммуникативной в сбытовую деятельность.

Что такое типичный рекламный посредник? Это лицо, предоставляющее возмездные посреднические услуги в доведении информации о товаре до потенциальных потребителей. Однако рекламный посредник не несёт абсолютно никакой ответственности перед заказчиком за эффективность рекламы. Не случайно до сих пор эффективность рекламы оценивается в количестве просмотров, охвате аудиторий, частоте показов и т.д. – в чём угодно, только не в показателях отражающих изменения показателей продаж заказчика в результате рекламных мероприятий.[13] Такая ответственность рекламщикам не нужна.

Так оно было, пока Интернет не трансформировался из канала коммуникации в канал сбыта продукции. Основная причина превращения заключается в том, что поставщикам не нужны рекламные и торговые посредники. Поставщикам нужны продажи. Электронная коммерция привела к появлению нового гибридного вида посредников, одновременно выполняющих рекламные и торговые функции. Однако, поскольку рекламная деятельность вторична по отношению к сбытовой, то она просто растворилась в ней.

Новые посредники в обмен на свои услуги вместе с торговой скидкой получают право продавать товары поставщика в сети Интернет, не имея их в наличии.[14] Они не создают товарных запасов у себя и не создают оборотных издержек у поставщика. Они конкурируют между собой, продвигая (рекламируя) товары поставщика в Интернете. Их аудитория не имеет географических ограничений, а предложения концентрируются на целевых аудиториях. Именно такие посредники олицетворяют собой сегодня электронные каналы сбыта в Интернет-маркетинге.

Поставщики получают в лице Интернет-посредников не только торговую инфраструктуру, инструмент ускорения оборота и источник информации о рыночной конъюнктуре, но и бесплатное продвижение (рекламу). За такое продвижение они платят торговой скидкой. Однако, в отличие от традиционной рекламы, оплата посреднических услуг находится в жесткой зависимости от показателей продаж, а показатели продаж зависят от разворотливости посредников.

Это принципиально важная тенденция в Интернет-маркетинге. Электронная коммерция ведёт к усилению интеграции между торговыми партнёрами, что выражается не только в делегировании полномочий по продаже товара контрагентам, но и в делегировании ответственности за эту продажу. В традиционном маркетинге такое было невозможно.

3. *Взаимодействие с покупателями.* Изменения в отношениях продавцов и покупателей, связанные с Интернет-маркетингом, обусловлены коренным изменением сущности взаимоотношений между продавцом и покупателем. Через Интернет, не только розничный торговец, но и производитель способен «дотянуться» до каждого покупателя. Например, через оказание сервисных услуг после регистрации на сайте и заполнения анкеты.

Традиционная теория маркетинга выделяет два вида маркетинга, отличающихся диаметрально противоположными подходами к организации маркетинговой деятельности.[15] Эти подходы определяют принципы и механизмы взаимодействия с потенциальными покупателями.

Первый вид, *потребительский маркетинг*, отличается отсутствием у покупателей должной информации о реальном качестве товара. Потребители ориентируются не на товар, а на сложившийся стереотип восприятия товара. Отсюда следует примат методов и форм продвижения, ассоциированных в основном с рекламой и PR.

---

[13] См. напр.: Уэллс У., Бернет Дж., Мориарти С. Реклама: принципы и практика. – СПб.: Питер Ком, 2008. – С. 163-170.
[14] Дойль П. Маркетинг, ориентированный на стоимость. – С. 431-432.
[15] Подробнее см.: Калужский М.Л. Практический маркетинг. – С. 33.

Второй вид, *промышленный маркетинг*, отличается тем, что покупатели обладают доскональными знаниями если не о самом товаре, то об особенностях его использования. Здесь конкурентное преимущество определяется уровнем технологического совершенства товара, а основным методом продвижения являются прямые продажи.

В традиционном маркетинге в обоих случаях речь в первую очередь шла о маркетинговых коммуникациях. Однако в Интернет-маркетинге коммуникации перестали играть определяющую роль по следующим причинам:

а) для индивидуальных потребителей подробная информация о товаре и его применении (в т.ч. негативная) стала общедоступной, а рекламное навязывание товара утратило свою эффективность;

б) у промышленных потребителей появилась возможность быстро получать конкурентные предложения и дополнительную информацию со всего рынка, что сделало ненужным визиты торговых представителей.

В традиционной торговой цепочке не только покупатели, но и производитель (поставщик) имел очень ограниченные возможности для сбора маркетинговой информации о состоянии потребительского спроса. Розничные торговцы не были заинтересованы в сборе подобной информации для производителя конкретного товара. В их ассортименте находились тысячи наименований товара, и они физически были не в состоянии собирать маркетинговую информацию о рынке и конкурентах для каждого поставщика.

Указанную проблему можно было частично решить через систему авторизованного дилерства, эксклюзивные скидки и т.д. В этом случае наиболее типичные торговцы в обмен на особые условия поставок предоставляют поставщику информацию о потребителях и конкурентах. Однако всё равно адекватность и оперативность информации о рынке, получаемой поставщиком от контрагентов, находилась в обратной зависимости к длине сбытовых каналов.

Интернет-маркетинг кардинально изменил ситуацию в пользу производителей и поставщиков. Несмотря на то, что товары продаются конечным потребителям через посредников, длина сбытовых каналов значительно сократилась, а производители получили полный контроль над ними. Даже если товар продаёт торговый посредник, а отгружает логистический посредник, процесс сбыта организует и получает всю информацию о продажах поставщик.

**Элемент II. Ценовая политика в Интернет-маркетинге**. Ценообразование и ценовая политика в электронной коммерции складывается в условиях, приближенных к условиям идеальной конкуренции. Ни один поставщик не может ограничить доступ покупателей к информации о ценах и конкурентных предложениях. Ни один посредник не способен монополизировать рынок и диктовать условия поставщику. Мало того, покупатели свободно обмениваются информацией о товарах между собой на специализированных форумах и в социальных сетях.[16]

Вместе с тем, естественная экономия на трансакционных издержках даёт Интернет-продавцам существенные преимущества по цене в сравнении с традиционной торговлей. Именно это преимущество является главным стимулом для покупателей к совершению Интернет-покупок.[17] Мало того, вовлечённость производителей в электронную коммерцию коренным образом меняет не только ценовую политику, но сам подход к ценообразованию. Вектор усилий виртуализующихся компаний направляется из сферы продаж (на которую влиять сложно) в сферу дальнейшего снижения издержек и получения за счёт этого новых ценовых преимуществ (на которую влиять легко и просто).

В результате конкуренция переходит из сферы борьбы за рынок в сферу адаптации к потребностям рынка. Парадокс ситуации в том, что борьба с конкурентами уступает место их сотрудничеству в деле сокращения не только трансакционных, но и любых иных издер-жек. Если на рынке присутствуют десять конкурентов, но только трое из них сотрудничают, объединяясь для сокращения издержек и решения совместных задач, то именно они объективно будут более конкурентоспособны.

---

[16] Дойль П. Маркетинг, ориентированный на стоимость. – С. 22-23.
[17] Интернет в России: Состояние, тенденции и перспективы развития. 2012. Отчёт. – М.: ОАО «НИЦ «Экономика», 2011. – С. 81.

В качестве примера можно привести IT-компанию «Covisint», созданную в 2004 году автомобильными концернами «Ford», «General Motors», «Daimler Chrysler», «Nissan» и «Renault».[18] Основная цель проекта заключалась в том, чтобы сократить стоимость производства одного автомобиля «*на 1 тыс. долларов за счёт объединения поставщиков, ускорения проектирования и разработки новых моделей, оптимизации моделей и сокращения складских запасов*».[19]

Сегодня противостоять экспансии электронной коммерции могут только крупные ритейлеры, успешно осваивающие сегодня этот рынок. Однако их возможности Интернет-маркетинга ограничены их же спецификой. Ритейлеры торгуют имеющимся в наличии товаром. Несмотря на льготы поставщиков и огромные объёмы, там велики товарные остатки и значительны трансакционные издержки.

Обычно поставщики отгружают товар ритейлеру в долг. Они не заинтересованы в новых поставках для полной оплаты за предыдущие партии, даже если товар устарел или не пользуется спросом. Поэтому наибольшую конкурентоспособность ритейл сохраняет на рынке товаров повседневного спроса с длинным жизненным циклом.

Виртуальные компании, наоборот, торгуют чужим товаром, отгружаемым со склада продавца или со склада логистического посредника. У них всегда есть последние модели и модификации товара и никогда не бывает нереализованных остатков. Поставщик отгружает товары напрямую покупателям и получает полную предоплату. Поэтому в сфере высокотехнологичных товаров неповседневного спроса с коротким жизненным циклом виртуальные компании всегда оказываются вне конкуренции.

Вместе с тем, традиционные инструменты ценовой политики (скидки, бонусы, зачёты за покупку, ценовая дискриминация и т.д.) в современном Интернет-маркетинге также присутствуют. Однако главный козырь Интернет-маркетинга виртуальных компаний – это возможность конкурентного снижения цен за счёт сокращения трансакционных издержек.

Ценовая конкуренция товаров и брендов с переходом маркетинга в Интернет превратилась в конкуренцию трансакционных издержек. Пока традиционные компании задают своими издержками верхнюю границу цены товаров, виртуальные компании будут пользоваться преимуществами более низких цен за счёт более низких трансакционных издержек.

Такая стратегия кардинальным образом меняет сам процесс ценообразования в маркетинге. Прежде трансакционные и производственные издержки определяли базовую цену продавца в категории «при прочих равных». Они были примерно равны и одинаково доступны для всех участников рынка. Маркетинговая политика заключалась в позиционировании новых (или как бы новых) товаров на рынке по завышенным ценам с последующим извлечением прибыли из разницы между базовой ценой и ценой реализации.[20]

Интернет-маркетинг, наоборот, в качестве базовой цены использует цены традиционных производителей. При этом продавцы отталкиваются не от ценовых представлений покупателей, а от сложившихся цен в традиционной торговле (см. Рис. 1). Не случайно подавляющее число рекламных кампаний в Интернет-маркетинге строится на ценовых сопоставлениях с традиционной торговлей.

| **Традиционный маркетинг** | **Интернет-маркетинг** |
|---|---|

---

[18] Сайт компании «Covisint». – https://www.covisint.com.
[19] Ермошкин Н.Н., Тарасов А.А. Стратегия информационных технологий предприятия: Как Cisco Systems и ведущие компании мира используют Интернет Решения для Бизнеса. – М.: Моск. Гум. Ун-т, 2003. – С. 288.
[20] Подробнее см.: Калужский М.Л. Практический маркетинг. – С. 45.

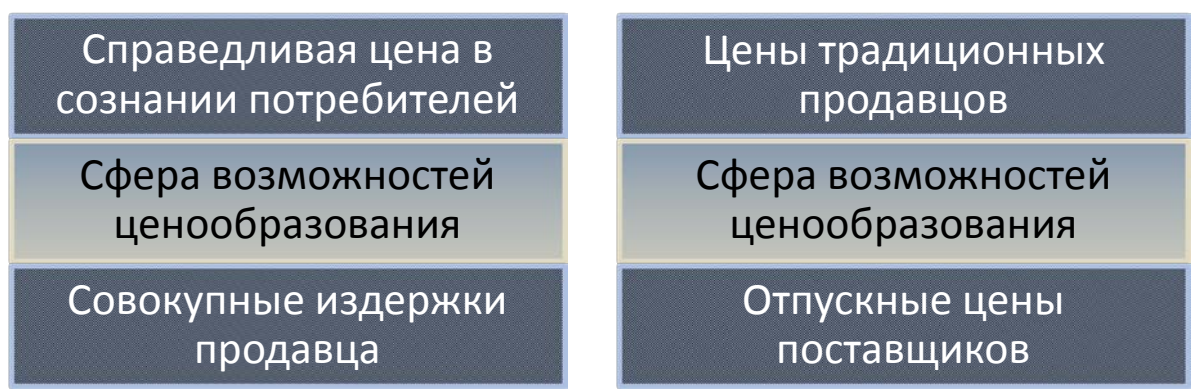

**Рис. 1. Особенности ценообразования в Интернет-маркетинге**

Некоторые зарубежные авторы указывают на ещё одну немаловажную сторону использования ценообразования в Интернет-маркетинге, рассматривая ценовую политику как «*рычаг, помогающий управлять спросом*».[21] Идея эта не нова. В традиционной теории маркетинга она реализуется через использование демаркетинга и ремаркетинга. Простейший пример: завышение цены с целью сократить спрос из-за невозможности его удовлетворить или, наоборот, занижение цены с целью стимулирования продаж.

В электронной торговле такой подход тоже наполнился новым содержанием. Цена здесь, так же как и в традиционном маркетинге, является инструментом обеспечения баланса спроса и предложения.

Однако в Интернете нельзя с такой же лёгкостью манипулировать покупателями. Это не похоже на единственный магазин в деревне, где только один выход – платить больше или ехать в другую деревню с непредсказуемым результатом. Здесь цены находятся возле минимума рентабельности, прибыль обретается за счёт роста объёмов продаж, а конкурентные предложения видны каждому покупателю. «Ценовая вилка» для манипулирования ценами тут невелика. Снизу её ограничивают отпускные цены поставщика, а сверху – высокий уровень конкуренции.

Поэтому в Интернет-маркетинге ценовые стратегии не играют столь важной роли, как в традиционном маркетинге. Скорее тут работает общеэкономическая закономерность, согласно которой цены способны завышать только монополисты (например, логистические посредники). Рядовые участники рынка способны конкурировать либо за счёт углубления кооперации, либо за счёт увеличения охвата рынка и объёма продаж.

Гораздо важнее другое. Ценовая политика в Интернет-маркетинге позволяет решать логистические проблемы, возникающие в процессе организации товародвижения (например, задержки в поставках и жалобы потребителей). Проблемы решаются через разделение сфер ответственности, тогда как покупателю предоставляется право выбора условий поставки и связанной с ними цены.

Если в традиционном маркетинге доминирует установление окончательной цены продавцом, то в Интернет-маркетинге продавец устанавливает лишь отпускную цену, а покупатель сам выбирает логистического посредника и связанные с ним риски. Таким образом, продавец несёт ответственность только за поставку товара логистическому посреднику, который самостоятельно вступает в правовые отношения с покупателем.

Например, покупая товар в сети Интернет, покупатель выбирает способ доставки: обычная почта, EMS или транспортная компания. Самый простой способ – обычная почта, но это медленно и высок риск порчи (воровства) товара. Скоростная почта «EMS» или «DHL» доставляет быстро, но дорого. Отказываясь от части прибыли в пользу логистических посредников, Интернет-продавцы попутно передают им бремя ответственности перед покупателями. Это ведёт к ещё большему снижению трансакционных издержек и позволяет сосредоточиться только на организации Интернет-продаж.

---

[21] Бергер Э.Дж. Цепи поставок: лучшие из лучших / Управление цепями поставок. Под ред. Дж.Л. Гатторны. – М.: Инфра-М, 2008. – С. 562.

Кроме того, виртуальные компании пользуются важнейшим фактором успеха в маркетинге – фактором времени.[22] Никто не способен так быстро принимать и реализовывать решения по ценам, как виртуальные компании.[23]

У виртуальных компаний нет проблем ни с обновлением ценников, ни с ведением бухгалтерского учёта. Они имеют дело с нематериальными (информационными) ресурсами и могут «*предлагать более быстрые платежи и решения на всех звеньях цепи поставок*».[24] Данное обстоятельство обусловлено тем, что хотя всё и сводится в конечном итоге к продаже реальных товаров, но область принятия решений по ценам располагается в виртуальном пространстве.

**Элемент III. Товарная политика в Интернет-маркетинге**. Согласно теории маркетинга товар – это «*всё, что может быть предложено для удовлетворения человеческих потребностей или нужд. ... Значение материальных продуктов состоит не столько в обладании ими, сколько в их способности удовлетворять определённые потребности*».[25]

Важное преимущество Интернет-технологий заключается в том, что они дают возможность перейти продавцам от выборочного сбора информации о реальном спросе на товары к получению полной информации о нём в автоматическом режиме «24-7-365». На основе этого преимущества в теории Интернет-маркетинга даже сформировалась новая концепция, получившая название «*концепция индивидуального маркетинга*».

Согласно данной концепции, наибольшая эффективность продаж достигается через предоставление потребителям индивидуализированных товаров и услуг более высокой ценности посредством интерактивных коммуникаций.[26] Речь идёт о том, что не только покупатель обладает ценностью для компании, но и компания обладает ценностью для покупателя, если она наилучшим образом удовлетворяет его запросы. Отказ от услуг такой компании ведёт к неоправданным потерям времени и усилий (трансакционным издержкам) покупателя на налаживание отношений с новым продавцом.

В традиционном маркетинге о трансакционных издержках покупателей никто даже не задумывался. Основными параметрами предложения там являются свойства товара и его цена – первые два элемента комплекса маркетинга. Это было обусловлено относительной недоступностью для покупателей информации обо всём спектре конкурентных предложениях. Покупатели часто не имели времени и возможности собирать информацию о товарах и сравнивать предложения на рынке.

В Интернет-маркетинге информацию практически обо всех конкурентных предложениях на рынке можно без труда найти в течение нескольких минут. Индивидуальные характеристики товара постепенно переходят в категорию «при прочих равных», а условия продаж приближаются к условиям совершенной конкуренции. Товар либо есть на рынке, либо его там нет. Поэтому товары в традиционном понимании в значительной степени утратили свою маркетинговую привлекательность для продавцов. Как отмечает по этому поводу П.Дойль: «*В современной экономике сфера услуг вдвое превосходит производственную, да и растёт значительно быстрее*».[27]

В результате произошла трансформация смыслового наполнения самого понятия «товар». Товаром стали именоваться логистические услуги по поиску, приобретению, доставке и оплате того, что ранее именовалось товаром.[28] Конкуренция в Интернет-маркетинге сместилась из сферы товарного производства в сферу логистического сопровождения продаж.

Поэтому своеобразной точкой отсчёта при формировании основ товарной политики в Интернет-маркетинге стал не общий маркетинг, а обладающий определенной спецификой

---

маркетинг услуг. В нём довольно давно существует понятие «внутреннего маркетинга». «*Цель внутреннего маркетинга*, – пишет Ф.Котлер, – *состоит в том, чтобы помочь служащим предоставить клиенту те товары или услуги, которыми он будет доволен*».[29]

Иначе говоря, внутренний маркетинг направлен внутрь фирмы с целью повышения степени её адекватности требованиям рынка. Отсюда следуют особенности распределения маркетинговых полномочий. Если в товарном маркетинге отдел маркетинга выполняет львиную долю маркетинговых функций, то в сфере услуг, наоборот, на отдел маркетинга приходится их минимальная часть.[30] Основная маркетинговая нагрузка в сфере услуг ложится на плечи сотрудников, непосредственно взаимодействующих с клиентами.

В Интернет-маркетинге товарная политика в сфере услуг наполняется новым, дополнительным содержанием. Вместо внутрифирменного персонала внутренний маркетинг переориентируется на выполняющих его функции логистических посредников. Происходит то, что П.Дойль назвал «*дезинтеграцией цепочек создания стоимости и реформированием отраслей*».[31] Участники процесса сетевой коммерции обладают абсолютной самостоятельностью по отношению друг к другу. Каждый из них предоставляет потребителям свою услугу в своей сфере деятельности.

Интеграция усилий участников цепей поставок под единым руководством в Интернет-маркетинге практически отсутствует. В этом заключается основа их конкурентоспособности, так как каждый субъект маркетинговой деятельности выходит за рамки отраслевой принадлежности, диверсифицируя, таким образом, свой рыночный потенциал. Например, один и тот же логистический посредник за счёт своей узкой специализации может одинаково успешно участвовать в продвижении медицинских услуг и совершенствовании автомобильных технологий.[32]

Все маркетинговые усилия в товарной политике Интернет-маркетинга направлены не на обеспечение уникальности торгового предложения (УТП), а на удовлетворение индивидуальных запросов потребителей. Как отмечает П.Дойль: «*Производство* [услуг и товаров] *на заказ, а не для заказов*».[33] Отсюда следует главная маркетинговая задача виртуальных компаний – найти конкурентную нишу на рынке и как можно дольше удерживать её. Такая ниша может быть связана либо с более или менее эксклюзивными условиями поставок, либо с конкурентным присутствием на узком целевом рынке. В общей теории маркетинга такой подход к организации маркетинга называется «сетевым подходом».[34]

Сетевой подход подразумевает, что каждый участник сети товародвижения обладает определенным статусом, понимаемым как роль, которую он играет по отношению к своим партнерам. Задача маркетинга в рамках сетевого подхода – приобрести выигрышный статус в сети, а затем укреплять и защищать свое положение. Основная идея заключается в том, что каждый участник сети зависит от контролируемых его партнерами ресурсов. Используя свой статус в сети, фирма получает доступ к их ресурсам. Поэтому рыночным ресурсом становится сам статус фирмы в сети. Наибольшее развитие сетевой подход получил в ритейлинговых сетях и в промышленном маркетинге.

Кардинальное отличие Интернет-маркетинга от традиционного маркетинга заключается ещё и в том, что здесь сетевой подход применяется не в сбытовой, а в товарной политике. Если в традиционном маркетинге сетевое положение приобретал доминирующий участник канала сбыта, то в Интернет-маркетинге уникальность его положения в сети определяет реакция конечных потребителей. Именно поэтому в комплексе Интернет-маркетинга компо-ненты «товар» (product) и сбыт (место, place) поменялись местами, но принадлежность сетевого подхода к третьему элементу комплекса маркетинга осталось неизменным.

---

[29] Котлер Ф., Боуэн Дж., Мейкенз Дж. Маркетинг. Гостеприимство. Туризм. – М.: Юнити-Дана, 2007. – С. 411.
[30] Там же. – С. 407.
[31] Дойль П. Маркетинг, ориентированный на стоимость. – С. 429.
[32] См. напр.: Сайт компании «Covisint». – https://www.covisint.com.
[33] Дойль П. Маркетинг, ориентированный на стоимость. – С. 443.
[34] Подробнее см.: Калужский М.Л. Практический маркетинг. – С. 27.

С позиций теории поведения потребителей, в электронной коммерции произошёл своеобразный перенос трансакционных издержек с производителей на покупателей. Они выбирают услугами каких логистических посредников будут пользоваться. «*Задача маркетинга заключается в том,* – отмечают М.Кристофер и Х.Пэк, – *чтобы найти пути увеличения ценности для покупателя, улучшив качество воспринимаемых преимуществ и/или уменьшив совокупные затраты на эксплуатацию*».[35]

Можно даже утверждать, что товарная политика в Интернет-маркетинге основной упор делает не столько на сокращение издержек участников продвижения, сколько на сокращение издержек покупателей. Благодаря этому с одной стороны обеспечивается большая свобода выбора для покупателей, а с другой стороны повышается привлекательность товарного предложения.

Суть нового подхода заключается в признании того, что покупателю не нужен огромный ассортимент товаров, который в Интернет-маркетинге вторичен. Покупателю нужен конкретный товар с искомыми параметрами, максимально удовлетворяющий его потребности. Задача продавца – не просто предложить максимальный ассортимент, но учесть индивидуальные запросы каждого конкретного покупателя. Только тогда покупатели, получив искомый товар с минимальными затратами сил и времени, вступят в долговременные отношения с продавцом, обеспечивая ему прибыль, сетевой статус и будущие объёмы продаж.

Поэтому обычный товар широкого доступа продать в Интернете достаточно сложно. Товар должен быть индивидуализирован под конкретного покупателя. Иначе говоря, по каким-то параметрам предложение товара в традиционной торговле должно не устраивать покупателя. Интернет-маркетинг продвигает не те товары, которых много, а те, которые востребованы. Покупатель приходит в Интернет, чтобы найти быстро и с минимальными трансакционными издержками именно «свой» товар. Тот, кто сумеет предложить каждому потенциальному покупателю «его» товар и станет лидером электронных продаж.

Индивидуализированность товарного предложения привела к смене общего вектора маркетинговой деятельности. Предлагающие товары логистические посредники превратились во владельцев информации о состоянии рыночного спроса, выступающих в отношениях с поставщиками от имени покупателей. Иначе говоря, вместо продажи покупателям товаров от имени поставщика, они стали обменивать на льготы и скидки поставщикам возможности закупок товаров на контролируемых ими целевых рынках.

**Элемент IV. Коммуникативная политика в Интернет-маркетинге**. Развитие сетевой коммерции существенно изменило характер маркетинговых коммуникаций. Из инструмента информационного воздействия на аудиторию коммуникации превратились в инструмент диалога с покупателями и контрагентами, а также инструмент принятия коллективных решений в Интернет-маркетинге. Благодаря Интернету они приобрели интерактивный характер.

Ф.Котлер пишет: «*Сегодня коммуникации рассматриваются как интерактивный диалог между компанией и её потребителями. Они осуществляются на этапах до совершения покупки, её совершения, потребления и после потребления*».[36] Продавцы получили возможность оперировать маркетинговой информацией не только о каждом товаре, но и о каждом покупателе, о каждой покупке этого покупателя. Это позволило объединить два взаимоисключающих подхода к построению маркетинговых коммуникаций: маркетинг сделок и маркетинг взаимоотношений.

*Маркетинг сделок* изначально доминировал в маркетинговой практике компаний в США и Японии из-за их ориентации на экспорт и большой длины торговых каналов. Он означает ориентацию на продажу унифицированных товаров массового спроса массовым покупателям.

---

[35] Кристофер М., Пэк Х. Маркетинговая логистика. – М.: ИД «Технологии», 2005. – С. 62.
[36] Котлер Ф., Боуэн Дж., Мейкенз Дж. Маркетинг. Гостеприимство. Туризм. – С. 624.

*Маркетинг взаимоотношений* доминировал в практике европейских компаний в силу их высоких издержек и ограниченности рынков сбыта. Он означает установление долговременных отношений с покупателями и индивидуализированный подход к их обслуживанию.[37]

Интернет-маркетинг сделал рынки безграничными и позволил индивидуализировать обслуживание большого количества клиентов. С одной стороны, в Интернете стали возможными коммуникации с целевыми аудиториями и с отдельными клиентами. С другой стороны, появилась возможность автоматизировать ведение маркетинговых баз данных и индивидуальные контакты с партнёрами и клиентами. В результате маркетинговые коммуникации в сети Интернет индивидуализировались, автоматизировались и обезличились одновременно.

Коммуникации стали первым элементом комплекса маркетинга, активно применяемым в Интернете даже тогда, когда электронная коммерция была в зачаточном состоянии. Вероятно, этим объясняется тот факт, что сегодня коммуникации уже во многом прошли путь эволюционного развития от инструмента распределения информации до набора унифицированных и автоматизированных функций.

Данное обстоятельство привело к формированию особого вида логистических посредников отвечающих за обеспечение информационного взаимодействия в виртуальной среде. Речь идёт об использовании в коммуникациях т.н. «*CRM-систем*» (*Customer Relationship Management*) – программных средств автоматизации взаимодействия с покупателями и контрагентами. CRM-системы используются сегодня для сбора и обработки маркетинговой информации, а также для ускорения обмена коммерческой информацией как внутри фирмы, так и между партнёрами.

CRM-система как модель взаимодействия с партнёрами в Интернет-маркетинге делает клиента основным объектом маркетингового анализа. В этом заключается принципиальное отличие CRM от логистических автоматизированных систем, где объектом анализа являются внутренние экономические параметры логистических потоков. Основное предназначение CRM-систем состоит в обеспечении процессов автоматизации электронных продаж и обслуживания клиентов в сети Интернет. CRM-системы автоматизируют сбор, обработку и анализ информации о контрагентах, поставщиках и потребителях, а также информационные потоки внутри компаний разной степени виртуальности.

В качестве примера можно привести CRM-систему германской компании «SAP AG». Ядром этой системы является клиентская база данных, на основе которой пользователи анализируют эффективность своих контактов с клиентами, связи клиентов, историю их покупок, контрактов и т.д. Как указано в справочном руководстве компании, CRM-система «*позволяет анализировать клиентов в различных разрезах и строить модели их поведения, в т.ч. на базе истории работы с ними*».[38] Используя возможности CRM-систем, продавец может заранее определить целевую аудиторию для маркетинговых коммуникаций, потенциал продаж, параметры предложения и другие характеристики коммуникаций. Остаётся только довести коммерческие предложения до тех покупателей целевого рынка, которые в них действительно нуждаются. Благо, то и другое Интернет-технологии позволяют сделать в автоматическом режиме, иногда даже без участия человека.

Технологизация касается всех компонентов коммуникаций в Интернет-маркетинге, превращая их в чисто технические функции по продвижению товара. Не случайно Ф.Котлер указывает, что электронная коммерция изменяет предназначение рекламы, которая носит в Интернет-маркетинге «*скорее информационный, нежели убеждающий*» характер.[39]

В электронной торговле, где все процессы автоматизированы и виртуализи-рованы, коммуникативные решения постепенно превращаются в набор опций на виртуальной панели управления покупками и продажами. При этом для пользователей эволюция Интернет-приложений идёт в обратном направлении: от сложного к простому. Сначала появляется по-стоянно усложняющийся механизм виртуального маркетингового решения проблемы. Затем на смену малодоступному для непрофессионалов механизму приходят простые в управлении и относительно дешёвые Интернет-сервисы.

---

[37] Подробнее см.: Калужский М.Л. Практический маркетинг. – С. 24-26.
[38] Решение SAP для управления взаимоотношениями с клиентами (SAP CRM). – Walldorf (Baden): SAP AG, 2008. –S. 8.
[39] Котлер Ф. Маркетинг менеджмент. – С. 783.

В качестве примера можно привести столь популярную некогда SEO (*Search Engine Optimization*) – поисковую оптимизацию сайтов, представляющую собой «*процесс достижения первых мест в результатах поиска в поисковых машинах по целевым для компании запросам*».[40] Ещё несколько лет назад электронная коммерция ассоциировалась с созданием Интернет-магазинов, посещаемость которых напрямую зависела от рейтинга информации о сайте в поисковых системах Yandex, Google, Rambler и др. До сих пор существует огромное количество виртуальных компаний, предлагающих за умеренную плату «поднять» рейтинг сайта.

Одновременно происходило совершенствование технологий расчёта рейтинга Интернет-ресурсов поисковыми системами. Процесс усложнился настолько, что разобраться в тонкостях и ухищрениях SEO-оптимизации человеку без глубоких знаний по этому предмету невозможно. Однако сегодня укрупнение Интернет-бизнеса и развитие технологий электронной коммерции позволяет продавцам осуществлять коммуникации с потребителями и без SEO ухищрений, создания сайтов или привлечения программистов.

Так, например, торговая площадка «Молоток» предлагает бесплатное создание и поддержание интернет-магазина с логотипом и уникальным адресом для юридических лиц.[41] Комиссия составляет от 2 до 5,5% с продажи товаров. Зарегистрированные продавцы получают бесплатные инструменты для управления продажами, а также доступ к аудитории с 500 000 потенциальными покупателями, ежедневно совершающими более 10 000 сделок. Зарегистрированные пользователи Молотка могут воспользоваться бесплатными опциями для организации промо-продаж товаров со скидками или платными опциями для отображения товаров на главной странице.

Зарегистрированным продавцам больше не нужно обманывать поисковые системы, продвигая свои сайты в Интернете. Любой начинающий предприниматель может без специальных знаний и усилий открыть собственный магазин на электронной торговой площадке, воспользовавшись всеми преимуществами электронной торговли.

В этом заключается ключевая тенденция развития маркетинговых Интернет-коммуникаций и сетевой торговли в целом. Повышение доступности электронной торговли одновременно обусловлено усложнением Интернет-технологий для разработчиков и упрощением маркетинговых решений для конечных пользователей. Коммуникации развиваются в Интернете не вертикально, а горизонтально.

Важной особенностью маркетинговых коммуникаций в Интернете является и то, что виртуализация электронных продаж ведёт не только к «*дезинтеграции цепочек создания стоимости*».[42] Покупатели, благодаря индивидуализации продаж, становятся полноправными участниками маркетинговых отношений. Они сами образуют виртуальные сообщества с торговыми посредниками, напрямую взаимодействуя с поставщиками товаров. Например, когда молодые матери объединяются в социальной сети для заказа партии детской одежды.

В результате размывается грань между внешним и внутренними маркетингом. Постоянные покупатели становятся частью виртуальной инфраструктуры сбыта и адресатом трансформирующихся внутрифирменных (внутренних) коммуникаций. Они сами начинают активно заниматься обратным маркетингом, направленным на посредников и продавца, влиять на принимаемые маркетинговые решения.

В результате сфера внутреннего маркетинга смещается в сферу маркетинговых коммуникаций. Классик американской теории маркетинга Ф.Котлер выделяет во внутреннем мар-кетинге сферы услуг четыре основных направления.[43] В Интернет-маркетинге эти направления видоизменились и институционализировались, но не утратили своей актуальности:

1. *Формирование культуры обслуживания клиентов* – трансформировалось в формирование норм и правил осуществления маркетинговых коммуникаций. Например, в случае невыполнения продавцом на Интернет-аукционе «eBay» правил продаж, покупатель вправе открыть диспут в платёжной системе «PayPal» и автоматически получить оплату обратно. Всё происходит в автоматическом режиме и коммуникации превращаются во внутренний сугубо технический процесс.

2. *Маркетинговый подход к управлению кадрами* – трансформируется в маркетинговые подходы к поиску и обеспечению взаимодействия с партнёрами, покупателями и логистическим посредниками. Интернет-компании всё чаще имеют виртуальную структуру, где альянсы создаются под проекты и заказы, а каждый участник отношений абсолютно самостоятелен и независим. В этой ситуации коммуникации направлены на обеспечение единства виртуальной организации и координацию усилий на достижение общих целей.

3. *Организация внутренних потоков маркетинговой информации* – в электронной торговле, как уже говорилось выше, стирается грань между внутренними и внешними потоками информации. Покупатель, включаясь в маркетинговые коммуникации продавца, становится частью его коммуникативной инфраструктуры, в которой потоки информации идут одновременно в двух направлениях.

4. *Система поощрений и признания среди служащих* – трансформируется в систему распределения прибыли между виртуальными партнёрами и в систему виртуального статуса. С одной стороны, рекрутирование сотрудников происходит в виртуальных профессиональных сообществах, где успехи и достижения каждого участника являются залогом деловых предложений в будущем.[44] С другой стороны, например, любая торговая площадка предусматривает систему индивидуального рейтинга для продавцов.

---

[40] Вирин Ф.Ю. Интернет-маркетинг. Полный сборник практических инструментов. – М.: Эксмо, 2010. – С. 94.
[41] Подробнее см.: Сайт электронной торговой площадки «Молоток.Ру». – http://molotok.ru/country_pages/168/0/shops/index.php#shops3
[42] Дойль П. Маркетинг, ориентированный на стоимость. – С. 429.

В результате маркетинговые коммуникации утрачивают свою первоначальную роль стимулятора покупательского спроса, превращаясь в технический инструмент информационного взаимодействия. Покупатель от этого только выигрывает, поскольку ему не приходится оплачивать затраты на дорогостоящие рекламные кампании и адекватность предоставляемой ему информации значительно возрастает.

Подводя итог, следует отметить, что, несмотря на реструктуризацию комплекса маркетинга в условиях электронной коммерции, сущность, цели, задачи и функции маркетинга не претерпели значительных изменений. Интернет-маркетинг превратился в самостоятельную и самодостаточную форму маркетинга, с присущими только её особенностями и механизмами реализации.

На повестке дня сегодня стоит дальнейшая институционализация сложившихся в рамках электронной коммерции отношений и формирование специальной теории Интернет-маркетинга, отражающей его институциональные особенности и приоритеты. Именно этот процесс, по всей видимости, будет являться ключевым направлением трансформации маркетинга в условиях бурного развития электронной коммерции в ближайшие годы.

---

[43] Котлер Ф., Боуэн Дж., Мейкенз Дж. Маркетинг. Гостеприимство. Туризм. – С. 411.
[44] См. напр.: Сайт социальной сети «Профессионалы.Ру». – http://professionali.ru.